\documentclass[aps,prb,twocolumn,showpacs]{revtex4-1}
\usepackage[dvips]{graphicx}
\usepackage{bm,amsmath}

\begin{document}

\title{Ring polymers with topological constraints}
\author{Yani Zhao}
\author{Franco Ferrari}
\email{ferrari@fermi.fiz.univ.szczecin.pl}
\affiliation{CASA* and Institute of Physics, University of Szczecin,
  Szczecin, Poland} 
\date{\today}
\begin{abstract}
In the first part of this work a summary is provided of some recent
experiments and theoretical results which are relevant in the
research of systems of polymer rings in nontrivial topological
conformations. Next, some  advances in  modeling the 
behavior of single polymer knots are presented.
The numerical simulations are performed with the help of the
Wang-Landau Monte Carlo algorithm. 
To sample the polymer conformation a set of random transformations
called pivot moves is used.  
The crucial problem  of preserving the topology of the knots after
each move is tackled with the help 
of two new techniques which are briefly explained. 
As an application, the results of an investigation of the effects of
topology on the thermal
properties of polymer knots is reported.
In the end, original results are discussed
 concerning the use of parallelized codes to study 
 polymers knots composed by a large number of segments
within the Wang-Landau approach. 
\end{abstract}
\maketitle
\clearpage

\section{Introduction}
Polymer knots and links (see Ref.~\cite{bookonknottheory} for a
definition of knots  and links) have been actively studied since the discovery
of interlocked polymer rings in DNA in 1961 by Frisch and Wasserman
\cite{Frisch1961}. After then, linked DNA molecules were identified in
HeLa cell in 1967 \cite{Hudson1967}. It has also been found that the
probability that open circular DNA molecules link with supercoiled
molecules is quite large \cite{Rybenkov1997}. 
Also in the production of artificial knots and catenanes
there has been a fast progress. The first artificial polymer knot, a
trefoil, has been synthetized in 
1989 \cite{Dietrich-Buchecker1989}.  

More recently, experiments have shown that bacterial DNA often occurs in
the form of knots that are sometimes heavily  linked together. For
example, DNA molecules extracted from tailless mutants of phage P4 vir
1 del22 are highly knotted (95\%) \cite{Arsuaga2005}. 
The mitochondrial DNA of Trypanosomes and related parasitic protozoa
consists of networks of thousands of topologically interlocked DNA
rings \cite{trypanosomes}. This abundance of knot and links in cells 
occurs because the probability of knotting increases with the degree
of confinement \cite{Micheletti2008}. Besides, a new mechanism of link
formation in DNA has been detected \cite{Elbaz2012}. 


Due to the progress of technology, the thermal 
and mechanical properties of polymer knots and links can now be
studied experimentally.  
The mechanical properties of single polymer knots have been
investigated for almost two decades with the help of optical tweezers  
and atomic force microscope tips.
It has been discovered in this way that polymer strands containing
knots are much more breakable under tension than straight filaments
\cite{Arai1999}.  
Moreover, since a few years also the collective behavior of many
polymer rings is accessible to experiments. 
The thermal properties of macromolecules forming knots and links may
be analyzed using calorimetric techniques. For instance, in polymer
fibers the presence of knots is revealed by irregularities in the
thermogram \cite{Steinmann2013}.  
In artificial polymer materials,  
the existence of knots and links  affects dramatically the
viscoelastic properties of polymers.  
In \cite{Kapnistos2008}  
melts composed by entangled polymer rings of an unprecedented purity ,
i. e. containing only a small fraction of linear chains, have been
obtained. 
It has been found 
by observing the power law stress relaxation of
these melts that they have a much lower viscosity than those
containing linear polymers, of about one order of magnitude less. What it
is interesting, is that even a very low amount of linear open chains
inside the melt is able 
to introduce relevant influence in the viscosity.
 For that reason, polymer knots and links in artificial materials can
 be used to fine tune the elastic behavior of the produced materials
 in industrial applications. The rheological properties of polymer
 melts of rings are actively studied in connection with their
 significant implications to our understanding of polymer dynamics.
 The mechanical properties of systems containing a few polymer rings
 entangled together under the tensions exerted by external forces
 play a relevant role also in the physics of DNA.  
More experimental results will come in the future thanks to more and
more refined techniques of synthesis and  separation of knotted
polymers. Already now, the formation of
polymer knots  and the effects of the topology in the intramolecular
interactions can be investigated experimentally, see e. g.\cite{Ohta2012}.  
  
The experiments mentioned above allow the comparison between the
observations and the predictions coming from theoretical models. For
this reason, they have attracted the interest of the community of
theoretical physicists. Theoretical models in turn provide a
microscopic understanding of what is observed and can point out the
directions of the future experimental research.
Many approaches to what can well be called the problem of topological
entanglement in polymer physics have been proposed. As a matter of
fact, polymers are already complex systems by themselves. When they
are further knotted and linked together, to this complexity one should
also add the topological complexity. This makes  the
treatment of systems containing polymer knots and links very hard.
The reward for solving this problem and achieving a better
understanding of polymer melts and materials composed by topologically
entangled polymer rings will be huge. The formation of links and knots
can result in fact in important effects in the physical behavior of
polymer materials, see Ref.~\cite{Orlandini2007} for a review
and further bibliography of this subject.

In describing the topological entanglement of polymer rings, it is
possible to take advantage of the progresses made in the previous
century in the classification of knots and links. In particular,
one should mention in this respect the construction of powerful knot
and link 
invariants like the
 polynomials of Alexander, HOMFLY and Conway or
  the invariants of Arf-Casson, Vassiliev--Kontsevich and several others. 
Following the seminal
work of Witten \cite{Witten1989} it has been possible to
derive expressions of knot and link invariants using a particular class
of field theories that are denominated topological.
To convince oneself that the abstract methods of knot theory really
matter in practical applications, it is sufficient to mention the
example
of the
DNA recombination procedure, during which
 the topology of DNA can be changed. The changes are performed by
particular enzymes called topoisomerases. The action of these
proteins cannot be observed directly, but it may be analyzed by
the methods of knot theory, that have been indeed successfully
applied in order to classify the effects of the topoisomerases
\cite{Sumner1996}.  

In polymer physics there has been always a nice interplay between experiments
and theoretical approaches, both numerical and analytical. Perhaps
the most important example of this is provided by the works of de Gennes and
coworkers \cite{degennes,cloizeaux}, that have led to a satisfactory
understanding of the behavior of polymers in a solution
thanks to the use of
renormalization group techniques.
A similar interplay occurs also in the research on polymer rings
subjected to topological entanglement.
The formation of knots in polymer
systems is probably the most well studied subject of the statistical
mechanics of polymer knots. It can be tackled analytically by
means of renormalization group methods. Nowadays there is a good
agreement between analytical and numerical estimations of how the
probability of formation of a knot of a given type scales with the
length of the polymer \cite{rensburg}. 
Numerical simulations on this subject have been started already in the
mid-seventies with the pioneering
works~\cite{Vologodski1974,Vologodski1975,Frank-Kamenetskii1975}.  
Analytical methods are also able to predict the various scaling laws
that characterize the asymptotic behavior of observables like the
gyration radius. Moreover, links between pairs of polymers
 can be analytically modeled
by using as the knot invariant
that takes into account the topology
the Gauss linking number. The model can be case in the form of
a  Ginzburg-Landau field
theory which is similar to those appearing in the physics of
critical systems. Its main characteristic is that the
scalar fields creating and annihilating the monomers of the polymer
trajectories are coupled with an
Abelian BF model. The latter is a topological field theory and describes
the "reaction forces" due to the presence of the topological
constraints. 
These constraints are necessary because, physically, two polymer
trajectories cannot penetrate themselves unless the temperature is so
high that the polymer melts down to single monomers or there are
enzymes like the topoisomerases that allow the opening and the
successive gluing back of the trajectories.
It is intriguing the fact that BF models are 
invariant under parity and time reversal transformations and have been
 used for this reason as effective field theories in the description
 of high $T_C$ superconductors and topological insulators. 
With the help of the topological Ginzburg--Landau model mentioned
above it has been
possible to formulate
concrete predictions on the behavior of linked polymers. First of all,
it has been shown that
the presence of the topological constraints on the polymer rings does
not affect their critical behavior \cite{FFIL2}. Nevertheless, it
affects the excluding volume interactions between the monomers by
weakening them \cite{FFIL2}. The results of \cite{FFIL2} are valid in the
approximation in which the monomer density is high and almost constant
apart from small fluctuations. Let us note that attractive forces
associated to topological constraints have effectively been observed
in an experiment \cite{levene}. Reviews on analytical methods can be
found in \cite{KholodenkoVilgisPhysRep} and in Kleinert's book
\cite{Kleinertbook}. 

Besides analytical calculations,  very reliable numerical
simulations allow to understand
the behavior of  polymer systems
observed during experiments. 
A wealth of publications
has been dedicated to 
the applications of polymer knots and links
in biology and biochemistry, like
for instance
\cite{Micheletti2008, Weber2006,Sulkowska2008,
  Sulkowska2009,Galera-Prat2010}.  
More and more complex problems are solved. Examples are the recent
advances in understanding the behavior of knotted proteins under
stretching \cite{Sulkowska2008, Sulkowska2009} and the breakability of
physical knots \cite{Pieranski2001}. Moreover, numerical simulations
have shown that in localized knots as those studied in
\cite{Arai1999}, the weakest points in the polymer strands are located
at the two points in which the knot starts and ends
\cite{Saitta1999}. Also numerical studies 
of the diffusion of polymer knots in gels have been able to reproduce
the experimental results \cite{Weber2006}. Let us remember that these
studies are relevant for the particular application of the phenomenon
of gel electrophoresis, that allows to extract polymer knots of given
types out of a mixture of polymer rings, e. g. DNA byproducts, having
different topological configurations. The thermal properties of
polymer knots in the stretched regime have been investigated very
recently in \cite{paea,yzff2013}. The latter works will be described
in the next Sections. 
Very recently, some important advances in the statics and
dynamics of polymer rings with or without entanglement have appeared
in the literature \cite{kremer2011,sommer2013}.
Numerical simulations are also important
to investigate phenomena that are hardly accessible by experiments.
For instance, despite the progress in understanding the viscoelastic
properties of melts of polymer rings mentioned above, it is still
difficult to isolate possible effects due to the presence in the melt
of knots or links.

The rest of the paper is organized as follows. A short review on
numerical approaches for
treating the topological constraints of polymer knots is contained in
Section \ref{aaa}. In Section \ref{sectionpaeatici}, we present two fast
techniques, namely the 
PAEA and TICI methods, which have been proposed in
\cite{paea,yzff2013} in order
to preserve the topology of
a polymer knot during the sampling procedure. The sampling
is performed with the help of the Wang-Landau Monte Carlo
algorithm, which is summarized in Section~\ref{sectionFd}. The PAEA
and TICI methods allow the sampling of a huge set 
of knot conformations as it is required in the investigation of the
thermal properties of unstretched polymer knots. 
Some of the results obtained with these methods are discussed in
Section \ref{bbb}. The original part of this work can be found in
Section \ref{sectionFd}, where the conclusions are drawn and some
further developments in the treatment of polymer knots with a large
number of segments are presented.
\section{Numerical approaches to the problem of topological
  entanglement in polymer physics}\label{aaa}
The key to study the thermal and mechanical properties
of polymer knots and links in both the analytical and
numerical approaches consists in being able to preserve the
initial topological 
configuration against the thermal fluctuations.
Let us note that throughout this work, the word configuration refers to a
particular topological state of a polymer knot.
The word conformation will instead denote the particular shape in the
space of the trajectory of a polymer knot in a given topological configuration.
Topological constraints may be
imposed 
on the possible knot conformations 
with the help of the so-called knot invariants (or link
invariants in the case of links).
Knot (link) invariants are mathematical
quantities whose values, when computed for a particular knot  (link),
do not  
change under any continuous deformation  of the knot (link), including
stretching of its spatial trajectory, but not for example cutting and gluing.
It should be kept in mind that there is no knot invariant that is able to
distinguish every knot uniquely. The same statement is true for link
invariants. 
The most common representations of knot and link  invariants are 
polynomials or  multiple contour integrals computed along the
physical trajectories of the polymers. Let us recall at this point
that these trajectories are usually
approximated by continuous curves following Edwards'
approach\cite{edwards}.
This approximation is particularly suitable for analytical
models. Numerically, the trajectories are discretized and become
mechanical systems of beads connected together by segments.
Some of the polynomial invariants, like the HOMFLY polynomials
mentioned before, 
are very powerful in detecting different topological configurations.
Recently, the A-polynomials and super A-polynomials have been
isolated in the amplitudes of topological string
theories\cite{Apolynomials}. These polynomials are able to
distinguish knots and links from their mirror reflections,
a feature that the HOMFLY polynomials do not possess.
Let us note that, so far, it has been impossible to fix the
topological constraints in
analytical models 
of topologically 
entangled polymers based on the Edward approach with the help of
polynomial invariants. The problem is that the coefficients of the
polynomials characterizing 
this kind of invariants 
cannot be easily related to 
the physical trajectories of the polymers. 
Only 
the invariants  expressed in the form of multiple contour integrals
have been successfully applied up to now.
This is the case of the Gauss linking number, which has been exploited
to describe polymer systems with topological constraints
imposed on  the links between pairs of polymer rings
\cite{FFIL,FFIL2,FFIL3}. Links in which three or four rings are
entangled may be described using Milnor type
invariants\cite{FFJMP2003,FFNOVA}. 
Unfortunately, there are no such simple invariants like the Gauss
linking number or the Milnor invariants that can be used to
distinguish the topology of a knot.  
The situation is different in numerical simulations, in which mainly
knot invariants in the polynomial form are considered, like the
already mentioned Alexander polynomials 
or the HOMFLY polynomials, see for instance
\cite{Vologodski1974,michalke}. While the Alexander polynomials are
not very powerful in detecting different knot topologies, their
numerical evaluation is fast. The more refined HOMFLY polynomials
require much more cpu-time to be computed. From this point on we will
concentrate on numerical simulations of single polymer knots.

In order to treat the statistical mechanics of knotted polymers two
main strategies can be devised.  
One strategy exploits self-avoiding random walks (SAW's)
\cite{probability1,probability2,probability3} on a lattice. A ring,
possibly in a nontrivial topological configuration, is
formed when the trajectory of the SAW intersects itself for
the first time. 
With this procedure, after considering many SAW's, it is  possible to
produce 
a statistically relevant amount of
polymer knots. 
The
probability $p_N({\cal K})$ of generating on a simple cubic lattice a
rooted lattice 
polygon with $N$ segments 
and a given topological configuration ${\cal K}$ scales is very well
known. It has been determined
analytically and checked numerically by several authors, see for
example  \cite{work1,work2}. For large values of $N$, the expression of
$p_N({\cal K})$ is given by:
\begin{equation}
p_N({\cal K})\sim N^{\alpha_S-2+N_{\cal K}}\mu^N\label{knotprob}
\end{equation}
where the parameters $\mu$ and $\alpha_S$ are the called the growth
constant and 
the entropic exponent respectively. $N_{\cal K}=1$ for prime knots. 
Of course, the type of the knot generated
after the SAW intersects itself is a priori unknown.
To determine it, knot invariants should be used.

The other strategy consists in
starting from a seed configuration of the polymer system
with a given topology. A statistically relevant set of different conformations
of the system is then achieved applying on it random transformations. 
These transformations should satisfy a few requirements.
First of all, they must be ergodic,
so that all conformations 
can be accessible. For instance, the pivot moves proposed in
\cite{Madras1990} have been proved to
be ergodic in the case of rings if their topological state is not relevant
 \cite{Madras1990,yzff2013}.  Basically, this means that, starting from
 a ring in an 
 arbitrary conformation and in an arbitrary topological configuration,
 after applying to it a finite number of pivot moves it is always possible
to arrive to a given seed conformation.
 However,  the final conformation and the intermediate ones are not
 constrained to have the same topological configuration of the initial
 ring.
In the case in which  the pivot moves are not allowed to destroy the initial
topology of the knot, which is relevant in the present context, there
is no rigorous proof of their ergodicity, but only numerical
evidences\cite{paea,yzff2013}. 

To preserve the initial topological configuration of a knot after
many random transformations, several different methods may be applied,
which can 
be based on knot invariants or not. 
The fastest way to avoid unwanted changes of topology makes use of
 random transformations that,
by construction, do not modify the topological configuration of the
polymer knot. This is for instance the case of the BFACF elementary moves 
introduced in Refs.~\cite{work3,work4}. 
In the literature \cite{work6}, the
ergodicity of the BFACF algorithm has been rigorously proven.
It has been shown in \cite{work7} that a Generalized Atmospheric
Sampling \cite{articleonGAS}
implementation of the BFACF algorithm is able to sample
the conformations of a trefoil knot consisting of lattice polygons
containing up to thousands of edges. 
The techniques based on the BFACF moves are
sampling the trajectories in the grand-canonical ensemble. 
This implies that the length of the knot may change, but  the
average
length can
be fine-tuned with an appropriate choice of the chemical potential in
such a way that most frequently trajectories of a given length are obtained.
There exist
also topology preserving random transformations that work directly in
the canonical ensemble, thus keeping fixed the length of the knot. An
example is  
provided by 
the pull moves of Ref.~\cite{pullmovesmainref}, which have been
applied in the case of polymer knots 
in \cite{swetnam}. 
The main problem of the transformations that automatically keep fixed
 the topological configurations of the system is that they
change only small portions of the knot trajectory. Especially for long
polymers, this leads to slow equilibration times and also increases
the time necessary for sampling the random conformations.
A compromise is to allow somewhat larger transformations, but
always not so large that a local analysis
near the
transformed element of the trajectory 
becomes insufficient to  detect potential topology
alterations. The first method of this kind,
 which is able to preserve the topology of knots on a simple cubic
 lattice by
forbidding the bond-crossings resulting from pivot moves, has been
proposed in 2012 \cite{paea}. The details of this approach
will be described in the next Section. A
method that is similar in spirit, but is valid for off-lattice
simulations, can also be found in the literature \cite{Narros}. The
idea of the algorithm of \cite{Narros} is to decompose the
collective pivot move, i. e. a move
involving more than one monomer after a random transformation, into  
successive elementary moves.  
After each elementary move, 
one considers the triangle whose vertices are given by
the new position of the 
moved monomer, it’s original position and the position of one of the adjacent
monomers.  
The topology
of the knot is preserved if there is no segment
composing the knot that
crosses the area spanned by such a triangle. The trial conformation is
accepted after all elementary moves are performed
and no bond-crossing takes place.
\section{The PAEA and TICI methods}\label{sectionpaeatici}
\begin{figure*}
\begin{center}
{
\includegraphics[width=0.5\textwidth]{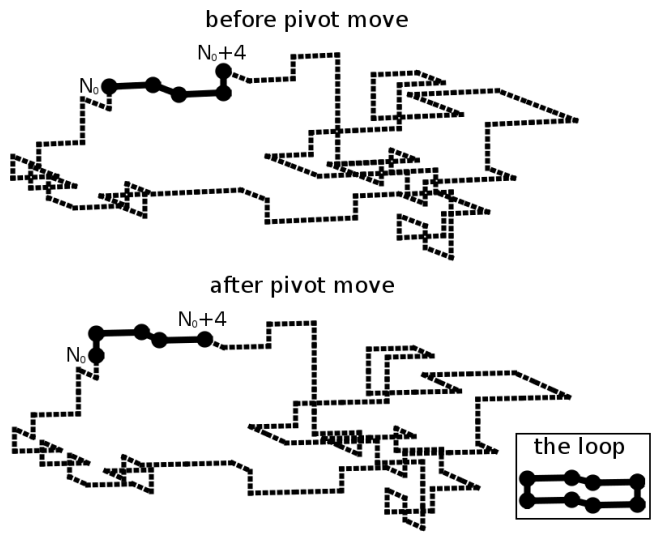}} 
\caption{\label{caseN=4a} The upper picture shows
a knot conformation ${\cal T}_X$ obtained 
  after applying to a seed conformation with
  the trefoil topology a number
  $X=4$ of pivot transformations. Each pivot transformation changed
  an element of the knot containing a number $K=4$ of segments. The
  lower picture shows the new knot 
  conformation ${\cal T}_{X+1}$ obtained after applying to ${\cal
    T}_X$ a trial pivot move.  The
small loop in the box appearing in the lower part of the picture in the right
  $\Delta{\cal T}_X$  is the difference between ${\cal
    T}_{X+1}$ 
and ${\cal T}_X$. The contour of $\Delta{\cal T}_X$ consists of the four
  segments from $N_0$ to $N_0+4$
belonging to  ${\cal T}_X$ which have been chosen for
  the pivot transformation and of the four segments 
also from $N_0$ to $N_0+4$
of  ${\cal
    T}_{X+1}$ obtained as a result of that transformation.} 
\end{center} 
\end{figure*}

\begin{figure*}
\begin{center}
{
\includegraphics[width=0.5\textwidth]{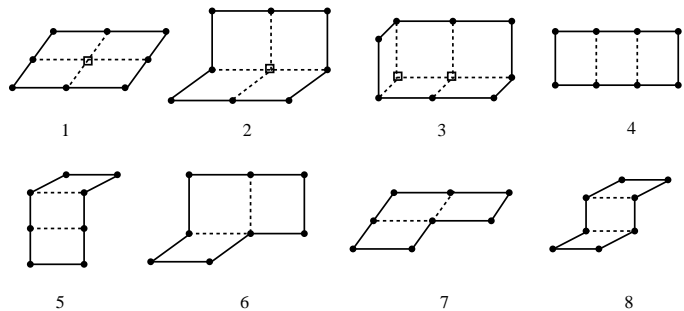}}
\caption{\label{smallloopgeometries} In this figure all possible
  conformations of the small loop $\Gamma$
up to possible rotations and reflections  
 have
  been listed for $K=4$. Arbitrary surfaces spanned by these loops
  have been drawn. The internal points have been denoted with squares.
}
\end{center} 
\end{figure*}

In this Section we restrict ourselves to numerical investigations on
the thermal properties of
single polymer rings with different knot types.
We choose the strategy of starting from a given seed conformation
of the knot to be studied and then acting on it with random
transformations in order 
to sample the set of all its possible conformations.
The case of the thermal properties of polymer knots under stretching
has been already studied a few years ago in \cite{swetnam}. Here we
treat unstretched polymers following Refs.~\cite{paea,
yzff2013}.
The problem of dealing with unstretched polymers is that they admit
much more conformations than stretched ones. Indeed,
for unstretched polymers containing a large number of
segments,  
an enormous number of conformations needs to be generated
to obtain a satisfactory statistics. 
If additionally the topological configuration of the polymer ring
needs to be preserved, it is very important to
 develop powerful and time-saving methods to perform random
 transformations of its trajectory without violating the topological
 constraints.
  To this purpose, two new techniques have been developed in
  Refs.~\cite{paea} and \cite{yzff2013}. 
The first one, which is valid on a simple cubic lattice, is the
so-called Pivot Algorithm and Excluded Area (PAEA) method
\cite{paea}.  The polymer is realized as an ensemble of beads, called
hereafter the monomers, connected together by segments of unitary length.
The PAEA method  uses as random transformations
the pivot moves of \cite{Madras1990}. 
These random transformations are applied on a randomly chosen element
of the knot trajectory starting from the $N_0-$th monomer and
containing a number $K$ of contiguous segments. Of course $N_0$ varies
randomly within the set of integers $1,\ldots,N$.
The strategy of
the PAEA method is based on the fact that the difference between the
old and new 
conformations after each pivot move results in a closed loop $\Gamma$ (or, if
$K$ is large enough, in a
set of closed loops). In figure \ref{caseN=4a} we show as an
example the loop formed after a pivot move on a trefoil knot $\cal T$.
Around the loop $\Gamma$ we span an
arbitrary surface having $\Gamma$ as its border. A trial pivot move is
rejected if the trajectory of the old knot conformation 
 crosses this surface or its border. It is easy to check that this is
 a necessary, 
 but not sufficient,
 condition for the trial pivot transformation to change the topology
 of the knot. Irrespective of the fact that the topology has been
 really changed or not, the trial pivot move is rejected if the
 surface or its border are crossed at least once.
This combination of 
pivot moves and the criterion of the excluded area, from which
originates the name of the algorithm, provides an
efficient and very fast way to preserve the topology that can be applied to
any knot configuration, independently of its complexity.  The time for
evaluating if the trial pivot moves has destroyed the original
topology or not scales as $\tau\sim N$.
The main disadvantage of the PAEA method
is that large pivot transformations are not easy to be implemented.
To understand why, we remind that
a pivot move involving
$K$ segments will result in a closed loop (or loops) $\Gamma$ counting
$2K$ segments. When 
$K=4$, one can easily classify all possible closed
loops with $2K=8$ segments that may arise after such moves and construct
appropriate surfaces having these loops as borders. The result are
 the eight loops
displayed in Fig.~\ref{smallloopgeometries}.
Let us note that loops 2--8 have no internal points that can be
intersected by the "old" trajectory of the knot, which is the
trajectory as it was before
the action of the random transformation. Thus, only the intersections
occurring at the borders of these loops must be checked. In the case
of loop 1, instead, besides the border there is also one internal
point that should be verified. If this point is intersected by the
"old" trajectory, then the topology of the knot obtained after the
random transformation has certainly been changed.
As $K$ increases, the number of 
closed loops becomes huge and the construction
of the surfaces having these loops as borders becomes a difficult task.
Up to now,
the PAEA method has been realized in the case of $K=4$ and $K=5$. Our
calculations based on the Wang-Landau Monte Carlo
algorithm\cite{wanglandau}, show that 
for polymers with  
$N=500$ segments, this is enough
to ensure the necessary ergodicity and a reliable statistics. For
longer polymers, 
the calculations may be finalized in a reasonable time only by means
of techniques of parallel computing, which will be discussed later in
Section~\ref{sectionFd}.
An alternative, or at least complementary way, consists in developing
methods that allow  to preserve the topology
for large random transformations, i.~e. involving large elements of the knot.
In fact, if the used random transformations affect only a small
portion of the
knot, the time for sampling increases as it was mentioned in the
previous Section. 
In order to detect possible topology changes due to large random
transformations, 
 knot 
invariants in the polynomial form are certainly very suitable.
As a matter of fact, apart from a few cases, they are quite powerful
in distinguishing the topology and can be applied to whatever knot
conformation, no matter how it has been changed after a random
transformation.
 However, excluding the Alexander polynomials, the calculation of more
 sophisticated polynomials is time consuming. Thus, 
in
\cite{yzff2013} it was explored the idea of  applying
 knot invariants in the form of multiple
contour integrals.  The contours are the knot trajectory itself or
elements of it.
From that idea it originated the TICI method, where TICI stands for
Topological Invariant 
in the form of Contour Integrals.
Examples of invariants of that kind
are abundant
in the physical and mathematical  literature, see for instance the
Arf-Casson invariant 
\cite{GMMknotinvariant}
 (equivalent to the Vassiliev invariant of degree
2\cite{otherauthormentioningVassilievknotinvariant})
or the
 Vassiliev-Kontsevich
invariants \cite{kontsevich}. 
The TICI method is based on the Vassiliev invariant of degree 2,
denoted hereafter $\varrho(C)$.

\begin{figure}
\centering
\includegraphics[width=0.5\textwidth]{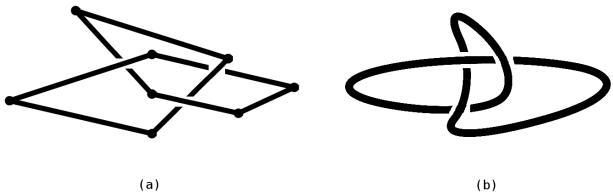}
\caption{A trefoil knot with eight segments defined off lattice
before and after the smoothing procedure.} \label{smoothing}
\end{figure}

Before \cite{yzff2013}, invariants in the
integral representation have never been applied in numerical simulations
of polymer physics. The only exception is the Gauss linking number, which has
been exploited
in numerical simulations\cite{ralf,Kremer1} of systems of linked
polymer pairs. 
The motivation of the little popularity of such invariants 
 is probably the fact that their
evaluation requires the computation of complicated
 multiple integrals. 
Let's us remark that the time $\tau$ needed to evaluate a multiple
integral with $l$ variables scales as $\tau\sim N^l$, where $N$ is the number
of segments composing the polymer knot. However, it has been noticed
in \cite{yzff2013} that the calculations can be sped up with the
help of a suitable Monte Carlo integration algorithm and of
parallelized codes. Besides, the
Vassiliev knot invariant of degree 2 on which the TICI method is based,
is
one of the simplest knot invariant whose integral representation is known. The
most time consuming integral to be computed is a quadruple one, so
that $\tau\sim 
N^4$. Restricting ourselves to random transformations in which 
the number of involved segments $K$
is much smaller
than the total number of segment $N$, this time can be
reduced\cite{yzff2013} to 
$\tau\sim N^3$. 
This feature makes the use of  $\varrho(C)$ competitive with respect
to the Alexander polynomials. In fact, using the fastest approximation,
the time required to evaluate the Alexander polynomial scales as
$\tau\sim (M-1)^3$, see 
\cite{koniarismuthukumar}. Here $M$ denotes the number of crossings
which is necessary to draw the knot after projecting it on an
arbitrary plane, see\cite{koniarismuthukumar} for more details. 
That approximation becomes not very precise when $M$ is large, a
situation which is common in polymers  confined in finite geometries.
A further advantage of invariants expressed as multiple contour
integrals with respect to polynomial invariants is their
portability. They can be computed on or off-lattice without any
modification. However, it should be mentioned the fact that their
evaluation requires a smoothing procedure which becomes relatively
complicated
in the general case of polymers defined
off-lattice \cite{ffyz2014}.
Fig.~\ref{smoothing} shows the effect of the smoothing procedure for an
off lattice trefoil knot.
Without the smoothing procedure,  the computation of the knot
invariant $\varrho(C)$ is affected by systematic errors due to the
presence of the sharp corners at the joints between contiguous
segments, but still  
it is possible to use it for the practical purpose of distinguishing
the topology of different knot configurations.
The comparison of the results of $\varrho(C)$ with and without the
smoothing procedure for several knots is shown in
Table~\ref{table1}\cite{ffyz2014}.   
\begin{table}[ht]
\centering
\begin{tabular}{|c| c| c| c| c|}
\hline
 knot type & $\varrho_a(C)$ & $\varrho_{sp}(C)$ & $\varrho_{ns}(C)$
 &$n_{sc}$\\ [0.5ex] 
\hline
$0_1$ &$-\frac{1}{12}$   &$-0.0839\pm 0.0332$ &$+0.5526\pm 0.0569$ &77\\
$3_1$ &$+\frac{23}{12}$  &$+1.9170\pm 0.0553$ &$+2.4781\pm 0.0465$ &68\\
$4_1$ &$-\frac{25}{12}$  &$-2.0847\pm 0.0533$ &$-1.5214\pm 0.0845$ &68\\ 
$5_1$ &$+\frac{71}{12}$  &$+5.9174\pm 0.0653$ &$+6.4523\pm 0.0845$ &65\\ 
$6_1$ &$-\frac{49}{12}$  &$-4.0856\pm 0.0723$ &$-3.5717\pm 0.1007$ &62\\ 
$7_1$ &$+\frac{143}{12}$ &$+11.9173\pm 0.0652$ &$+12.4258\pm 0.1217$ &62\\ 
$8_1$ &$-\frac{73}{12}$  &$-6.0822\pm 0.0529$ &$-5.6380\pm 0.0774$ &54\\ 
$9_1$ &$+\frac{239}{12}$ &$+19.9158\pm 0.0855$ &$+20.4041\pm 0.1579$ &59\\ [1ex]
\hline
\end{tabular}
\caption{The values of the Vassiliev knot invariant of degree 2
for the knots with Alexander-Briggs notation $0_1$, $3_1$, $4_1$,
$5_1$, $6_1$, $7_1$, $8_1$ and $9_1$. $\varrho_a(C)$ denotes the
analytical value of the knot invariant. $\varrho_{sp}(C)$ is the
results of the knot invariant with the smoothing procedure  
 described in \cite{ffyz2014}.
$\varrho_{ns}(C)$ is instead the value of the knot invariant
derived without the smoothing
procedure. The data of $\varrho_{sp}(C)$  
and  $\varrho_{ns}(C)$ have been
computed using the same number of samples, which varies depending on
the kind of knot. $n_{sc}$ is the number of sharp corners contained in
the knot before the smoothing procedure.}  
\label{table1}
\end{table} 
In writing Table~\ref{table1} we took advantage of the fact that
$\varrho(C)$ is related to the second
coefficient $a_2(C)$ of the Conway polynomials \cite{GMMknotinvariant}
by the following equation: 
\begin{equation}
a_2(C)=\frac 12 \left[\varrho(C)+ \frac{1}{12}\right]
\end{equation}
Since the coefficients $a_2(C)$ can be computed analytically for any type of
knots, the analytical values of $\varrho(C)$ are also known for any
given knot configuration. 
Compared with the PAEA method, 
the use of $\varrho(C)$ reduces
the number of samples necessary for the calculations of the averages
of the 
observables with the Wang-Landau Monte Carlo algorithm\cite{yzff2013}.
This reduction 
is probably due to the fact that with $\varrho(C)$ large pivot transformations
can be exploited, which are able to change relevant portions of the
knot. In this way, the exploration of the whole set of available
conformations becomes faster.
Despite the decreasing of the number of samples,
the computations still last in general
longer than those performed with the PAEA method, because the
expression of $\varrho(C)$ contains
quadruple integrals that should be evaluated numerically and this
is time consuming. 
Several tricks to reduce this time have been proposed in
Refs.~\cite{yzff2013,ffyz2014}. The most effective is the possibility
of reducing on a simple cubic lattice the number of segments $N$ by a
factor three.

\section{Thermal properties of polymer knots}\label{bbb}
As an application of the fast methods presented in the previous
Section, the thermal properties of several knots are computed using
the Wang-Landau 
algorithm \cite{wanglandau}. This has been done in Ref.~\cite{paea}
with the help of the PAEA method. The implementation of the TICI
method to the study of the statistical mechanics of polymer knots can
be found in Ref.~\cite{yzff2013}. The computational details of the
TICI method may be found in Ref.~\cite{ffyz2014}.
Here a brief account of these results will be provided.
Polymers are defined on a simple cubic lattice, with the monomers
located at the sites of the lattice. Very short-range forces
between the monomers are assumed. The related potential
is defined as follows: 
\begin{equation}
V_{IJ}=\left\{
\begin{array}{rl}
+\infty&\mbox{if $I=J$}\\
 \varepsilon&\mbox{if $d=|\vec {R}_I-\vec {R}_J|=1$ and
  $I\ne J\pm 
  1$}\\
0&\mbox{otherwise}
\end{array}
\right.\label{potdef}
\end{equation}
where $\varepsilon$ is the
interaction energy between pairs of non-bonded
monomers. The condition $\varepsilon< 0$ refers to the case of attractive
forces, while $\varepsilon>0$
characterizes the repulsive case.
Moreover, $\vec{R}_{I}$ denotes the
position vector of the $I$-th segment.

\begin{figure*}
\centering
\includegraphics[width=0.5\textwidth]{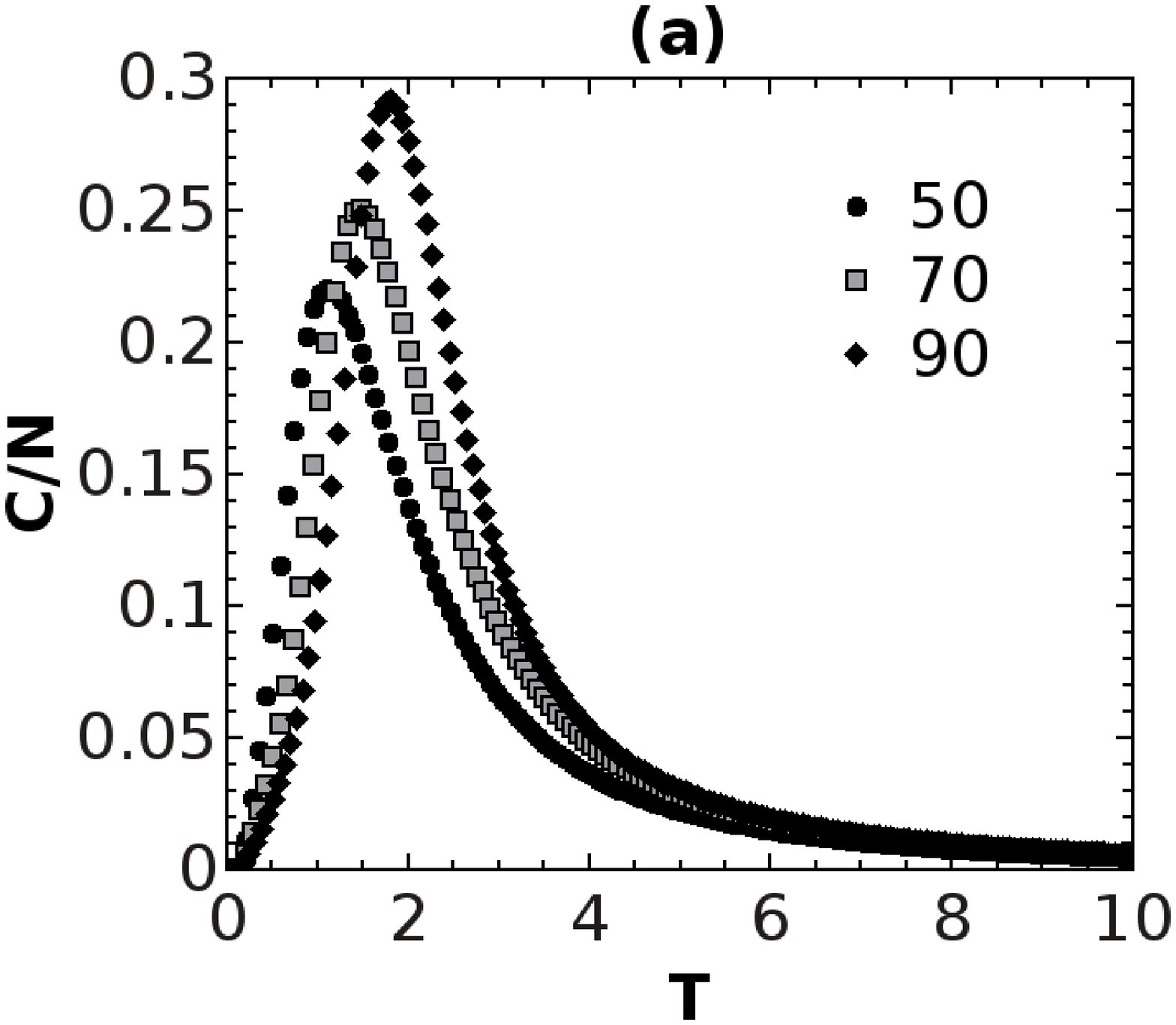}
\includegraphics[width=0.5\textwidth]{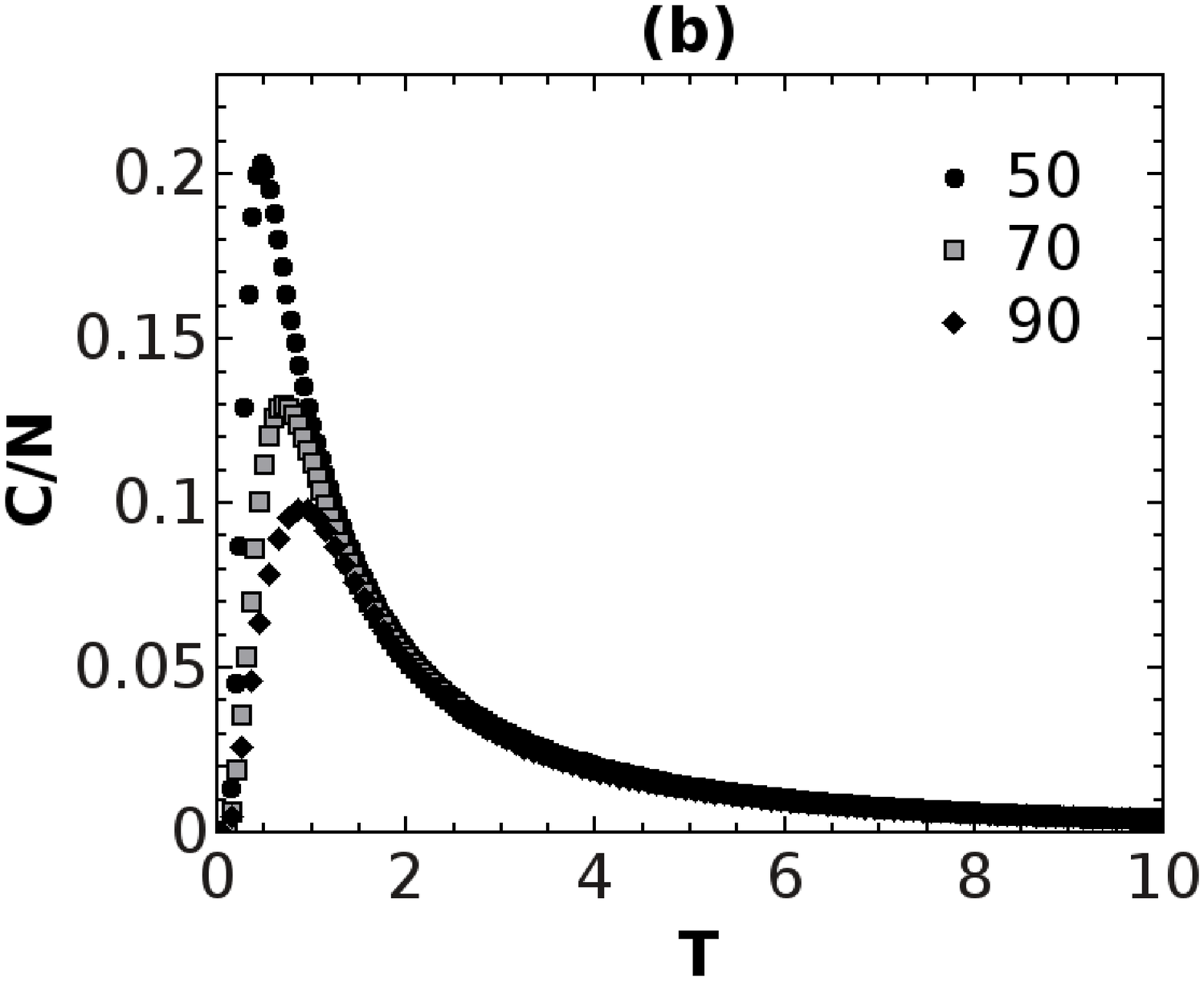}
\caption{\label{fig2}  The specific heat capacity (in units of
  $\varepsilon$) of the trefoil as a function of the normalized
  temperature $\mathbf T=\frac{T}{\varepsilon}$ in both (a) the
  attractive and (b) the repulsive cases. The number of segments can take the
  values $N=50$ (circles), $N=70$ (rectangles) and $N=90$ (diamonds).}  
\end{figure*}

\begin{figure*}
\centering
\includegraphics[width=0.5\textwidth]{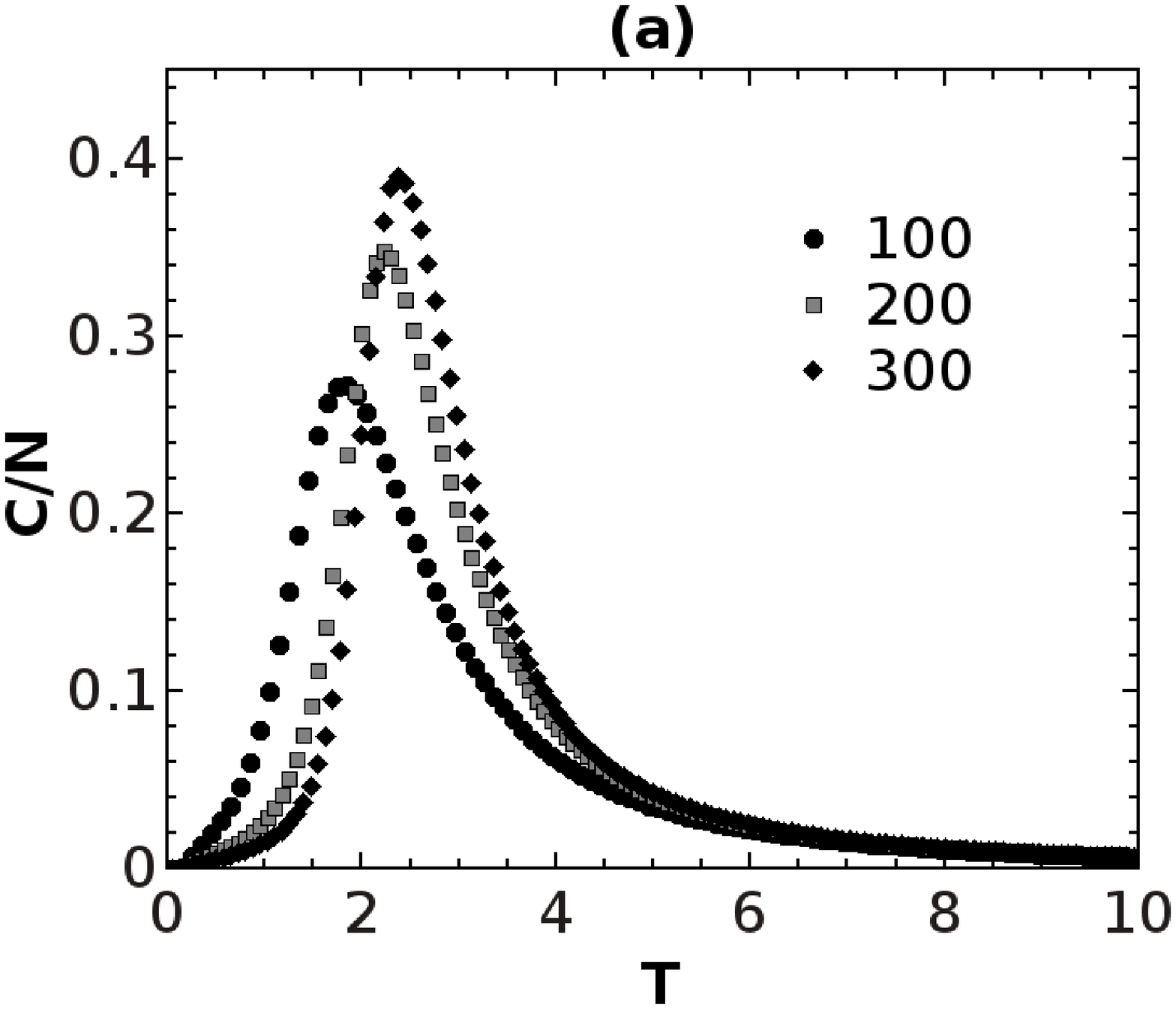}
\includegraphics[width=0.5\textwidth]{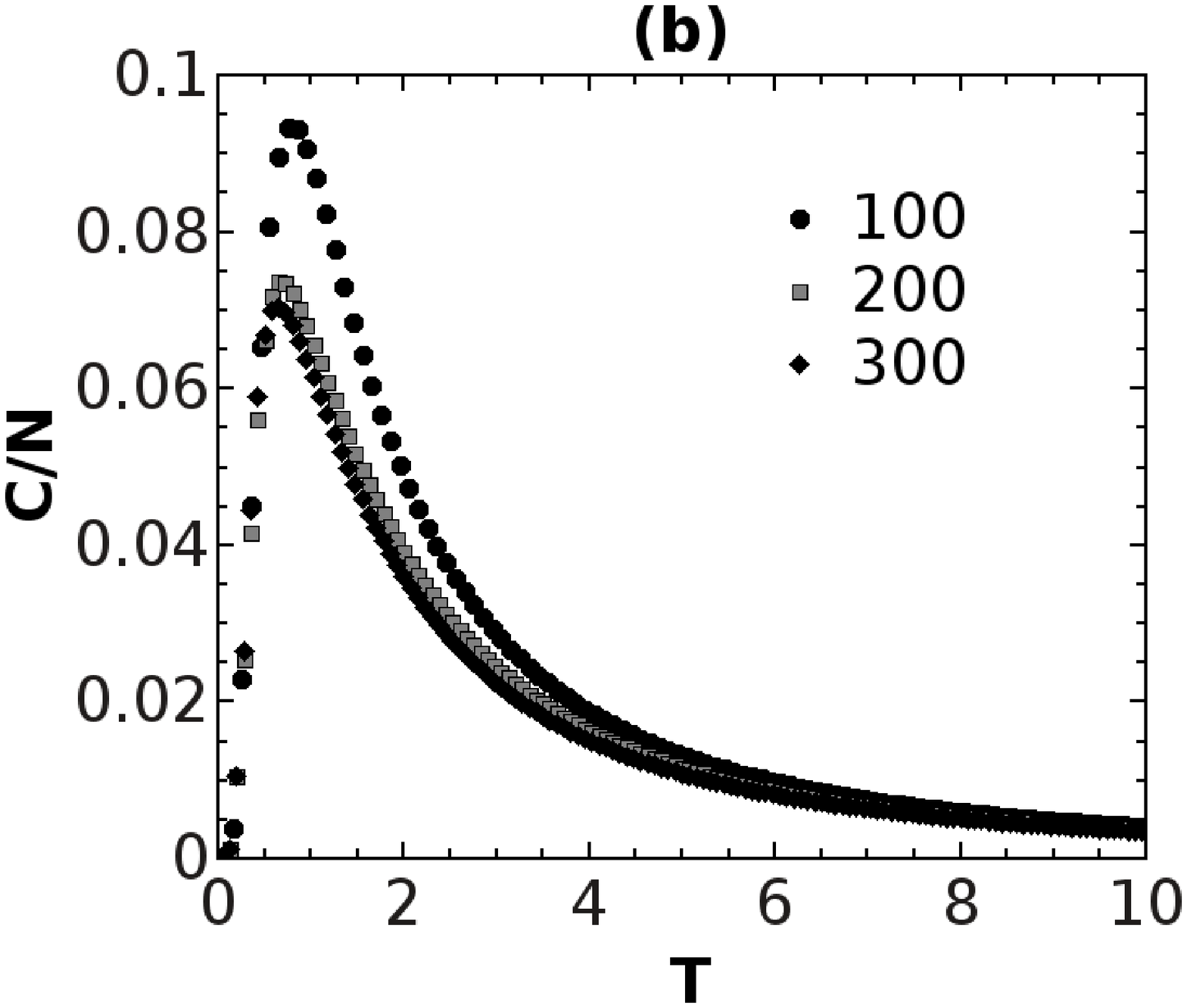}
\caption{\label{fig2-bis}  The specific heat capacity (in units of
  $\varepsilon$) of the trefoil as function of the normalized
  temperature $\mathbf T=\frac{T}{\varepsilon}$ in both (a) the
  attractive and (b) the repulsive cases. The number of polymer
  segments can take the 
  values $N=100$ (circles), $200$ (rectangles) and $300$ (diamonds).}  
\end{figure*}

In Refs.~\cite{paea} and \cite{yzff2013} the specific
energy, the specific heat capacity and the gyration radius of several
types of polymer knots have been computed.
In Fig.~\ref{fig2} we show the results of the computation
of the specific heat capacity obtained with the 
TICI method for a trefoil knot in both attractive (left panel) and repulsive
cases (right panel) \cite{yzff2013}.  The rings contain a relatively
small number of segments $N\le 90$, because the purpose of
\cite{yzff2013} was to study the influences of topology on the
thermal behavior of polymer knots. These influences are much more
marked if the polymers are short. This point will be discussed later
in further details.
%
We discuss here mainly the case in which the potential is attractive,
corresponding to
Fig.~\ref{fig2} (a).
The peak in the specific heat capacity is interpreted as a
pseudo phase
transition from a frozen crystallite state to an expanded state.
This is a pseudo phase transition because
  we are working with a
finite size system, far from the thermodynamic limit, as discussed in
Refs.~\cite{wuest,janke}.
In \cite{janke} it has been stressed that
such pseudo phase transitions
 will become more and more important, because they will soon be
observable
in real systems  thanks to the advances in
the construction of high resolution equipment.
The behavior of the specific heat capacity has been related with
the presence of a pseudo
phase transition by observing that the peak of the heat capacity at
the pseudo phase transition
grows more or less linearly with the increasing of the number of
segments $N$ as it is expected. Indeed, the peak of the
specific heat 
capacity remains at an almost constant height independently of the
value of $N$. This fact is evident also from Fig.~\ref{fig2-bis}~(a),
where polymer knots with $N=100,200$ and $300$ are considered.
Another reason hinting that a pseudo phase transition is undergoing is
coming from the plot of the gyration radius, which during the
transition increases considerably (more than 50\%), see Ref.~\cite{yzff2013}.
The nature of the initial and final states has been decided by
examining directly the knot conformations. Before the transition, at
low temperatures, the knot exhibits a partially ordered structure
similar to that of a crystal, with defects which are probably related
to the topological constraints and the knotting.
Similar pseudo phase transitions 
from a frozen crystallite state to an expanded state
have also been detected
 in the case of a single polymer chain discussed
in~\cite{binder}. We see only one
peak, because we are dealing with very short-range interactions, 
in agreement with Ref.\cite{binder}, in which it was found that, if
the
range of the interactions is very short, 
then the open chain admits just two possible states, namely the crystallite and
the expanded coil ones. 
Let us notice  that pseudo phase transitions under stretching have been
already observed in knots, see \cite{swetnam}. 

The presence of sharp peaks in the heat capacities in the repulsive
case, see Figs.~\ref{fig2} (b) and \ref{fig2-bis} (b), has not a
straightforward 
interpretation like 
those occurring when the interactions are attractive. 
The data concerning the gyration radius, in fact, show only a modest
increase of this quantity in the range of  temperatures in which the supposed
pseudo phase transition is undergoing. Moreover, the temperature at
which the peak occurs is rather low and, actually, the height of the
peak seems to decrease with increasing numbers of segment $N$.
As argued in \cite{yzff2013}, 
the peak in the heat capacity is almost probably due to a lattice
artifact, related to the fact that, 
when the temperature is very low, the first energy state
$E_1=\varepsilon$ cannot be reached, because $kT<E_1$. So the system
stays in the ground level $E_0=0$ and only when $kT$ becomes big
enough, it jumps to the next states $E_1,E_2,\ldots$.
Let us remark that the behavior of the
thermal properties in the repulsive case
is in agreement with the previous results of Ref.~\cite{work8}, where
it has been studied the dependence on the ion concentration of the
specific energy and the gyration radius of a mixture of knotted and
unknotted polymer rings in a salty solution.
The comparison is made difficult by the fact that the systems and the
interactions discussed here and in \cite{work8} are
different. However, in the repulsive case discussed here, we expect that the
very short-range interactions become 
irrelevant when the temperature is high. Analogously, when the ion
strength is low, the polymer knot is immersed in a good solvent,
thus experiencing repulsive forces, which are fading away with the
increasing of the ion strength. As a consequence, we can qualitatively
compare the behavior of a polymer knot for increasing temperature and
the behavior exhibited by the knot for growing ion strength.

\begin{figure}
\begin{center}
\includegraphics[width=3in]{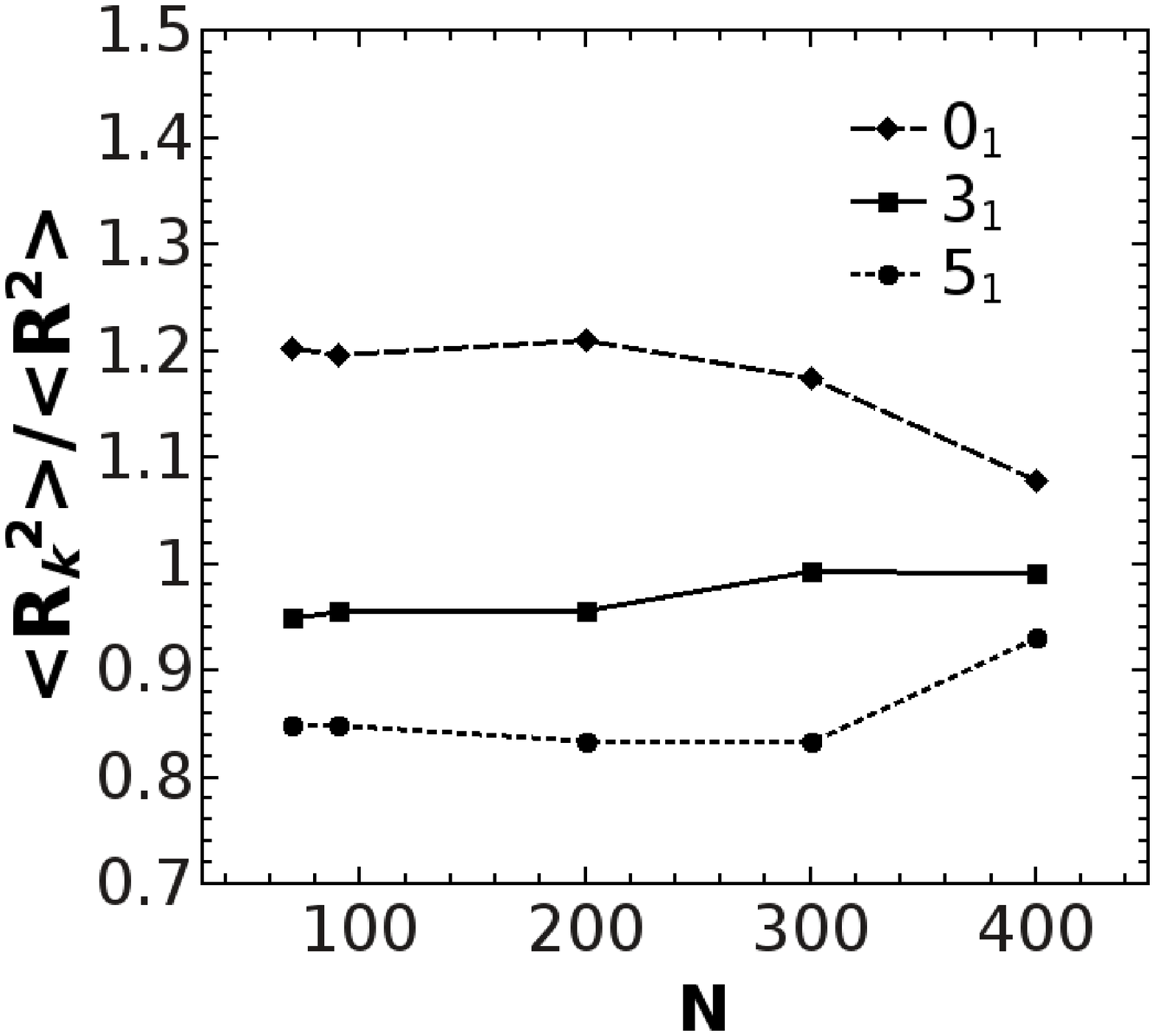}
\caption{\label{asymptbehav}
Plot of the normalized gyration radius
 $\frac{\langle R^2_{\cal K}\rangle}{\langle R^2\rangle}$
for knots of different topology ${\cal K}=0_1,3_1,5_1$ and of different
number of segments $N=70,90,200,300,400$. For increasing values of $N$, the 
normalized
gyration radii of different knots appear to converge to the same value
$1$.} 
\end{center}
\end{figure}

The topological effects on the thermal properties of knotted polymer
rings have been discussed in the paper \cite{yzff2013} by comparing
knots of different types but of the same length. It was shown
in\cite{yzff2013} 
that the topology of knots plays an important role when the size of the
polymer is small. Moreover, topology related effects disappear with
increasing polymer lengths.  
This fact can also be confirmed by the calculation of the gyration radii.
Following the work of \cite{deguchi3}, for instance, the values of the
normalized 
gyration radius $\frac{\langle R_{\cal K}^2\rangle}{\langle R^2\rangle}$ 
can be plotted as in
Fig.~\ref{asymptbehav} for different polymer lengths up to $N=400$.
Here $\langle R^2\rangle=\frac 13\sum_{\cal K}
\langle R_{\cal K}^2\rangle $ denotes the average of the gyration radius
of a closed polymer trajectory of fixed length
irrespective of its topological configuration $\cal K$.
The sum over ${\cal K}$ has been limited to the
knot types
${\cal K}=0_1,3_1,5_1$, which is the reason of the factor $\frac 13$ in the
expression of $\langle R^2\rangle$. 
Clearly, in the case in which there are no effects on the gyration
radius related to the fact
that the three knots $0_1,3_1,5_1$ are topologically different, we
would have that $\frac{\langle R^2_{\cal K}\rangle}{\langle R^2\rangle}=1$
for ${\cal K}=0_1,3_1,5_1$.
The  normalized gyration radius 
$\langle R_{\cal K}^2\rangle $
of each knot
${\cal K}$ with ${\cal K}=0_1,3_1,5_1$ has been computed using a set
of $50\times 10^6$ 
conformations.
Fig.~\ref{asymptbehav} 
suggests that the values of the normalized gyration radius
converge in the limit $N\longrightarrow\infty$, a fact that implies
that the dependence on the knot type
is disappearing with increasing number of segments $N$. 
\section{Further developments and conclusions}\label{sectionFd}
To preserve the topology of polymer knots 
during the sampling procedure needed in Monte Carlo simulations, two
algorithms, namely the PAEA and the TICI methods,
have been presented in Section.~\ref{sectionpaeatici}. The
performance of the PAEA method is independent of the complexity of the
knot and the time needed for accepting or rejecting a given conformation
after a random transformation grows linearly with the number of
segments composing the knot. This makes the PAEA method very fast. Its
disadvantage is the difficulty of its implementation for random
transformations involving more than $K=5$ segments.
On the other side, the TICI method is significantly slower, because
the time for computing the Vassiliev invariant of degree 2 scales with
$N$ as $\tau\sim N^4$. 
After several improvements, with the TICI method it is currently possible to
treat in a reasonable time the statistical properties of polymer knots
up to 400 segments. 
The advantage of this method is that large
portions of the knot may be changed randomly with a single
transformation. This speeds up the sampling of different
conformations.
Moreover, the TICI method works both on and off-lattice without
modifications and is easily generalizable to the case of links
composed by three linked rings.
As a matter of fact, the triple Milnor  invariant
describing the topological configurations of links among three rings
consists in a linear combination of terms that are very similar to the
Vassiliev invariant of degree 2 used in the TICI method.

Despite these progresses, the treatment of very long polymers or of
many polymers within the Wang-Landau algorithm becomes difficult when
$N\sim 1000$. 
For this reason, one should rely on parallelized codes. The
parallelization of the Wang-Landau algorithm has been discussed in
Ref.~\cite{yanlan} and, more recently, also in \cite{tholan}
The reader is addressed to these works for additional bibliography on
this subject. Here we would like to present two ways to implement the
parallelization of the Wang-Landau algorithm which are suitable for
polymer simulations.
Before doing that, however, a few words about the Wang-Landau
terminology is in order. 
The details are reported in the original article \cite{wanglandau}.
The goal of the Wang-Landau procedure is to
compute the so-called density of states $\phi_m$ for the energy
values $E_m$ in a given energy domain $D$.  In principle $D$ should
cover the whole energy spectrum. We assume for simplicity that the
various energy levels $E_m$ are labeled by positive integer numbers, so that
$m=1,2,\ldots$. 
$\phi_m$ represents  the number of  conformations having
energy $E_m$. Its expression is:
\begin{equation}
\phi_m=\sum_{X} \delta(H(X)-E_m)
\end{equation}
where $X$ is an arbitrary conformation with fixed topological
configuration and $H(X)$ is the Hamiltonian.
$\phi_m$ plays the same role of the number of states in the
microcanonical ensemble. Its relation with the partition function $Z$
in the canonical ensemble is:
\begin{equation}
Z=\sum_{m} e^{-\beta E_m}\phi_m
\end{equation}
with $\beta$ being the Boltzmann factor.
The Wang-Landau algorithm computes the density of states
perturbatively in a finite number of steps. At the
$\nu-$th step,
the precision is determined by the so-called modification factor
$f_\nu$. Each $f_\nu$ is defined by the relation: $f_\nu=f_0^{\frac
  1{2^\nu}}$, where $f_0>1$ is the value of the modification factor at
the zeroth approximation. Usually, $f_0$ is chosen to be equal to $e$.
At the beginning, the density of states has the initial value
 $\phi_m=1$ for all energy levels. Successively,
$\phi_m$ is updated with the following procedure. At each approximation 
order $\nu$, different conformations of the knot are randomly
generated.
A conformation $X_{m'}$ of energy $E_{m'}$
obtained after a random transformation of a
previous conformation $X_m$ of energy $E_m$, is accepted
unconditionally if the
transition probability:
\begin{equation}
p(X_m\longrightarrow X_{m'})=\min\left\{
1,\frac{\phi_m}{\phi_{m'}}
\right\}\label{cdf}
\end{equation}
is equal to one. Otherwise, a random number $\eta$ such that
$0<\eta<1$ is generated and the new conformation $X_{m'}$ is accepted
if $p(X_m\longrightarrow X_{m'})>\eta$. In all other cases $X_{m'}$ is rejected.
If $X_{m'}$ has been accepted, then the density of states and the energy
histogram $h_\nu(E_m)$ are updated as follows:
$\phi_{m'}\longrightarrow f_\nu\phi_{m'}$ and
$h_\nu(E_{m'})\longrightarrow h_\nu(E_{m'})+1$. The next approximation
level $\nu+1$ starts when the
histogram $h_\nu(E_m)$ of the conformations at the $\nu-$th
order becomes flat within a precision of $\pm 20\%$.

A possible parallelization strategy for performing the sampling
necessary in the Wang-Landau algorithm consists in splitting the
task into a number $J$ of threads $t_1,\ldots,t_J$ . All threads explore
simultaneously 
conformations in the whole energy spectrum. 
The threads work in cycles. 
At the beginning of a cycle, the global density of states
$\phi_m'$ is the same for all threads.
During the cycle, each thread samples a set of $\ell$
conformations and updates independently of the others the density of
states. The value of $\ell$ depends on the number of threads.
At the end of the cycle, every thread $t_s$, with $1\le s\le J$,
has updated the  density of states $\phi_m^{\prime}$ by a factor
$\Delta\phi_{m,s}^\prime$, The new density of states according to the
$s-$th thread will be thus given by:
$\phi_{m,s}^\prime=\phi_m^{\prime}\Delta\phi_{m,s}^\prime$.  
The new global density of states $\phi_m^{\prime\prime}$
is obtained by averaging over all the results coming from the $J$ threads:
\begin{equation}
\phi_m^{\prime\prime}=\sqrt[J]{\prod_{s=1}^J\phi_{m,s}^\prime}\label{update}
\end{equation}
At this point, a new cycle starts and the density of states
$\phi_m^{\prime\prime}$ is updated separately a number $J$ of times by the
threads. 
After sampling an additional set of $\ell$
conformations, the next corrections $\Delta\phi_{m,s}^{\prime\prime}$
to $\phi_m^{\prime\prime}$ are computed and the global density of
states $\phi_m^{\prime\prime\prime}$ may be obtained using the
geometrical averaging of Eq.~(\ref{update}).
The
procedure continues until the 
energy histogram becomes flat, after which the next approximation
level 
is started.
With this method, the threads are allowed to sample the set of all
possible conformations 
independently, but only for the limited number of samples $\ell$
generated during a cycle. When
the cycle ends, the information collected by the different threads gets
"exchanged" by the averaging process of Eq.~(\ref{update}). This procedure
avoids the situation in which the threads update the
density of states in a totally
independent way. Large scale simulations show that 
in this case there can be difficulties in estimating
 the correct density of states
$\phi_m$, see Ref.~\cite{tholan} on that point. 

Using the parallelization techniques described above, together with the
PAEA method to detect topology changes of the knot after a random
transformation, 
it is possible to 
study polymer knots subjected to very short interactions like those of
Eq.~(\ref{potdef}) in a reasonable time of weeks and with a satisfactory
level of approximation if the number of segments in the knot is
of the order $N\sim 1000$ or less. For a polymer with $N=1000$, the
approximation level $\nu=15$ is reached within the time of one month
using $J=12$ threads.
For greater values of $N$,
the number of  threads should be increased. Thus, to prevent possible
problems related to the simultaneous running of a huge amount of threads,
following \cite{tholan} it is reasonable to 
 adopt a strategy in which the energy domain is split into
many different regions $D_1,\ldots,D_L$. 
In each domain $D_j$, $j=1,\ldots,L$, the energies $E_m$ are limited to the
interval $[E^i_{min},E^i_{max}]$.
The energy values in these
regions partly overlap, i. e. the set $D^j\cup
D^{j+1}=[E^{j+1}_{min},E^j_{max}]$ is not empty. 
The goal is to evaluate separately the partial densities of states
$\phi^1_m,\phi_m^2,\ldots,\phi^L_m$ in the regions $D^1,D^2,\ldots,D^L$
respectively. As we will see, the global density of states  $\phi_m$
can be reconstructed if all the partial densities of states are known.
Of course, as mentioned in \cite{yanlan} and
\cite{tholan}, with the splitting of the total energy domain in
subdomains some statistically relevant set of
conformations may be ignored. For polymers, this is particularly true
in the case of very compact conformations. Indeed, if during the
sampling a class of very compact conformations is obtained, sometimes
it could be necessary to unpack the knot to some extent and then pack it
 again in another way in order to reach a class of even more compact
 conformations.
When the knot gets unpacked, the distances between the monomers
increase in the average, a fact that is connected with a 
change of the total potential energy of the system. If the sampling is
restricted to a particular energy region, it may well be happen that the energy
value that should be attained to allow the  unpacking of the knot lies
outside that region. This simple example shows how all  conformations
that may be obtained 
only by unpacking and repacking the knot starting from a given seed
conformation could become not accessible after the splitting of the
energy domain in many regions. 
To circumvent this problem, we choose a somewhat different approach from
that of \cite{tholan}, that will be called here the splitting
method. 
We assume now that we are just interested in computing the density of
states $\phi^i_m$ in the particular interval of energies
$D_i=[E^i_{min},E^i_{max}]$ for some value of $i$, with  $1\le i\le L$.
The basic idea is that to speed up the calculations,
avoiding to have to consider the whole energy spectrum,
it is not necessary to restrict the sampling only to the
region $D_i$.
It  is sufficient to limit the
time spent by the code in analyzing the other regions. To this
purpose, the 
exploration of the regions $D_j$ with $j\ne i$ can be penalized by
increasing appropriately the modification factor $f_\nu$ outside
$D_i$. Let $f_{i,\nu}^j$ be the modification factor 
that will be used in the interval $[E^j_{min},E^j_{max}]$, $j=1,\ldots,L$,
to evaluate the density of states $\phi_m^i$.
We define the $f_{i,\nu}^j$'s as follows: 
\begin{equation}
f_{i,\nu}^j=\left\{
\begin{array}{ccc}
f_{0}^{\frac1{2^\nu}}|E^{j}_{max}-E^i_{min}|^{\alpha_j}&\mbox{for}&j<i\\
f_{0}^{\frac1{2^\nu}}|E^{j}_{min}-E^{i}_{max}|^{\alpha_j}&\mbox{for}&j>i\\
f_{0}^{\frac1{2^\nu}}&\mbox{for}&j=i
\end{array}
\right.\label{modfactors}
\end{equation}
where the $\alpha_j$'s are suitably chosen constants.
In other words, 
 in the energy region of interest $D_i$,
$f_{i,\nu}^i$ is the usual modification factor of the Wang-Landau
 algorithm at the approximation level $\nu$.
However, the $f_{i,\nu}^j$'s have been increased by factors that are
proportional to some power $\alpha_j$ of the minimal distance on the
energy axis between
the energy values in $D^i$ and  $D^j$.
A good choice of $\alpha_j$ is for instance $\alpha_j=1$ for $j\ne i$.
With the splitting method just outlined, the sampling is performed in
every energy region, but 
most of the time is spent to sample the conformations in the selected
region $D_i$. 
When the energy histogram becomes flat in the interval
$[E^i_{min},E^{i}_{max}]$, the next level of approximation in computing the
density of states $\phi^i_m$ is started.
The simulations show that also
configurations whose energy is not in the chosen region $D_i$ are
visited several times, but of course much less than the conformations
with energy in the range $[E^i_{min},E^{i}_{max}]$, because the transition
probability (\ref{cdf}) to all conformations outside $D_i$ is
suppressed by the choice of modification factors in
Eq.(\ref{modfactors}). Let's now consider the case of regions in which
 the monomer density is high. Exactly these conformations are
 difficult to be sampled as discussed earlier. Assuming that the
 interactions are 
 attractive, they correspond to regions in which the energy is very
 low. If, starting from a given seed conformation, the system is trapped
 during the sampling in a class of 
 conformations of very 
 low energy which is not statistically relevant, with the method
 explained before the knot is still able to slowly unpack 
 itself under the effect of random transformations reaching domains
 of higher energy. After some time is passed, a longer stay in these domains
becomes very unlikely due to the
transition probability  (\ref{cdf}) and the choice of modification
factors (\ref{modfactors}).
As a consequence, the conformations
 drift back toward the region of very low energy and the knot gets
 packed again. After a conformation $X_{low}$ with energy belonging to the
 selected region $D_i$ is reached, the
 system remains in that region for a long time.
The conformation $X_{low}$ becomes the new seed, starting from which a
large number of conformations in the selected energy region is sampled.
Of course, we expect that, especially
for long polymers, 
after the knot is unpacked and repacked
in this way several times, it attains seed conformations
$X_{low}^{(1)},X_{low}^{(2)},X_{low}^{(3)},\ldots $ that belong to a
class of conformations with energies in $D_i$ whose number is
overwhelmingly larger than that of more rare conformations in the same
energy range.

\begin{figure}
\begin{center}
\includegraphics[width=3in]{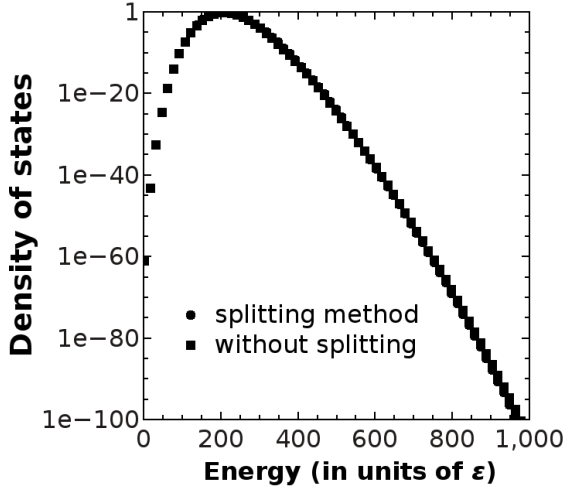}
\caption{\label{splitted51} Comparison of the densities of states for
  a knot $5_1$ with number of segments $N=1000$ computed with the
  splitting method 
  (dots) and without it (squares).}  
\end{center}
\end{figure}

In Fig.~\ref{splitted51} we compare
the densities of states for a knot $5_1$ with $N=1000$
segments computed with or without the splitting method. 
The total density of states $\phi_m$ has been obtained from the
partial densities of states $\phi_{i,m}$ derived from the splitting
method as follows. 
First, it is checked if
at the intersection between two contiguous energy domains $D_i\cap D_{i+1}$,
 the partial density of states in $D_{i+1}$ coincides with that of
 $D_i$. Let  $(\phi_{i,m},\phi_{i+1,m})$ and
 $(\phi_{i,m'},\phi_{i+1,m'})$  be two pairs of densities of states
 corresponding to the energies $E_m$ and $E_{m'}$ respectively.
$\phi_{i,m}$ and $\phi_{i+1,m}$, as well as $\phi_{i,m'}$ and
 $\phi_{i+1,m'}$, are the partial density of states resulting from the sampling
in the regions $D_i$ and $D_{i+1}$ respectively.
If both energy values $E_m$ and $E_{m'}$ belong to the intersection domain
$D_i\cap D_{i+1}$, then we have found from our simulations that the ratio
$\frac{\phi_{i,m}}{\phi_{i+1,m}}$ is almost equal to the ratio
$\frac{\phi_{i,m'}}{\phi_{i+1,m'}}$ within the required level of
approximation $\nu$. This means that two partial densities of states
calculated in contiguous energy regions $D_i$ and $D_{i+1}$ are
related together by a 
proportionality factor $C_{i,i+1}$ which can be defined as the geometric
average of all ratios $\frac{\phi_{i,m}}{\phi_{i+1,m}}$ on the
intersection $D_i\cap D_{i+1}$:
\begin{equation}
C_{i,i+1}=\sqrt[G]{\prod_{E_m\in  D_i\cap D_{i+1}}
\frac{\phi_{i,m}}
{\phi_{i+1,m}}}
\end{equation}
$G$ denotes here the number of energy values that are in common between the
regions $D_i$ and $D_{i+1}$.
Finally, the total density of states $\phi_m$ may be reconstructed 
from the partial densities of states $\phi_{i,m}$ as follows:
\begin{equation}
\phi_m=\left\{
\begin{array}{ccc}
\phi_{1,m}&\mbox{for}&E_m\in D_1\\
\phi_{2,m}C_{1,2}&\mbox{for}&E_m\in D_2\\
\vdots&&\\
\phi_{i,m}C_{i-1,i}&\mbox{for}&E_m\in D_i\\
\vdots&&
\end{array}
\right\}
\end{equation}
As it is possible to see from Fig.~\ref{splitted51}, the results of
the density of states computed with the splitting method coincide with
the results computed  by sampling the whole energy region.

Besides the study of the thermal properties of longer polymers, the
mechanical properties of knots can also be considered. Here we
restrict ourselves to the force-extension behavior of stretched polymer
knots. For that purpose, two different ensembles can be considered
\cite{threeensembles}: 
\begin{itemize}
\item Stress ensemble: In  this ensemble the tensile forces 
and their application points are known "thermodynamic"
parameters. The goal is to compute the 
resulting extension of the polymer at equilibrium.
An
  example 
in which the stress ensemble  has been applied
to study the statistical mechanics of single knotted  polymer rings
under stress on a
  simple cubic lattice, has been
  presented in \cite{swetnam}. 
\item Strain ensemble: In this case
the distance between two points of the knot is fixed and the average
forces at these points is evaluated.
\end{itemize}

Finally, in order to have a realistic physical model describing the
mechanical properties of a polymer knot under stretching, one should
simulate the 
stretching of the chemical bonds by allowing the segments to change
their length. This goal can be achieved by constructing polymers
using monomers which interact with their nearest neighbors by the
FENE potential \cite{kremerFENEpot}. With this set-up, off lattice
calculations become 
preferable. For the task of sampling different conformations while keeping
fixed their topology,
at least in the case of relatively short polymers the Vassiliev knot
invariant of degree 2 discussed in \cite{yzff2013} is very suitable,
because it may be easily applied to off lattice simulations. For 
polymers containing a large number of segments, new codes with a high
degree of parallelization or reliable 
techniques for performing the sampling in split energy intervals should be
developed. The splitting method outlined above seems a good candidate to
compute the density of states with the Wang-Landau algorithm using the
strategy of splitting the whole energy domain into many different
intervals.
It has been tested up to now for several knots with different number
of segments up to $N=1000$. Further
 investigations are necessary to assess its validity for larger values
 of $N$. 

\acknowledgments {The support of the Polish National Center of
  Science, scientific project No. N N202 326240, 
is gratefully acknowledged. The simulations reported in this work were
performed in part 
using the HPC cluster HAL9000 of the Computing Centre of the Faculty
of Mathematics 
and Physics at the University of Szczecin.}

\end{document}